\documentclass[aps,prl,reprint,superscriptaddress,amsmath,amsfonts,amssymb,longbibliography]{revtex4-2}

\usepackage{graphicx}
\usepackage{hyperref}
\usepackage{ragged2e}
\usepackage{soul,xcolor}
\usepackage{color}
\usepackage{dcolumn}
\usepackage{braket}
\usepackage{bm}
\usepackage[english]{babel}

\definecolor{mygreen}{RGB}{24, 119, 242}  

\begin{document}


\title{Fundamental Limits to Cat Code Qubits from Chaos-Assisted Tunneling}
\author{Lionel E. Mart\'inez}
\affiliation{Departamento de F\'isica ``J. J. Giambiagi'' and IFIBA, FCEyN,
Universidad de Buenos Aires, 1428 Buenos Aires, Argentina}
\affiliation{Department of Physics \& Astronomy, University of Southern California, Los Angeles, CA 90089, United States of America}

\author{Ignacio Garc\'ia-Mata}
\affiliation{Instituto de Investigaciones F\'isicas de Mar del Plata (IFIMAR),\\ Facultad de Ciencias Exactas y Naturales,
Universidad Nacional de Mar del Plata\\ \& CONICET, Funes 3350 (7600) Mar del Plata, Argentina}

\author{Diego A. Wisniacki}
\affiliation{Departamento de F\'isica ``J. J. Giambiagi'' and IFIBA, FCEyN,
Universidad de Buenos Aires, 1428 Buenos Aires, Argentina}

\date{\today}
\begin{abstract}  
We show that chaos-assisted tunneling (CAT) imposes an intrinsic limit to the protection of Kerr-cat qubits. In the static effective description, tunneling between the quasidegenerate cat states can be exponentially suppressed, ensuring long lifetimes. However, our Floquet analysis reveals that when the nonlinearities increase, chaotic states mediate tunneling between the cat states, producing large quasienergy splittings. We compute tunneling rates using both full quantum simulations and semiclassical WKB theory, finding quantitative agreement and confirming that the splittings are directly linked to chaos. These results provide the first evidence of CAT in the Kerr-cat qubit and demonstrate that chaos sets a fundamental bound on the coherence of dynamically protected superconducting qubits.
\end{abstract}
\maketitle
The realization of fault-tolerant quantum hardware relies on qubits with long coherence times and robust intrinsic error suppression. The Kerr-cat qubit (KCQ) \cite{Grimm2020, mirrahimi2014dynamically} encodes logical states in coherent-state superpositions stabilized by two-photon driving and Kerr nonlinearity. In the effective static picture, tunneling between the two quasidegenerate states can be parametrically suppressed, providing hardware-efficient protection against bit-flip errors \cite{venkatraman2024drive,refFrance}.
{Larger Kerr nonlinearities are actively pursued in next-generation
superconducting circuits, since the nonlinear timescale $1/K$ governs the
speed and selectivity of cat-based gate protocols and nonlinear control
operations \cite{Heeres2017,Eickbusch2022, Grimm2020,frattini2024observation,Lescanne2020,puri2017engineering,Frattini2018}. Exploring the strongly nonlinear regime is therefore essential
for assessing the ultimate performance limits of Kerr-cat qubits.}

This effective description, however, neglects the strongly nonlinear and periodically driven nature of the physical device. The actual implementation -- typically a superconducting Kerr oscillator with parametric drive -- can host classical chaos when nonlinearities increase. 
Recent studies have connected chaotic behavior in Kerr oscillators to delocalization phenomena \cite{GarciaMata2024effectiveversus}, to the destruction of excited-state quantum phase transitions \cite{cejnar2021excited,chavez2024driving,garciamata2025impact}, and to the onset of chaos in coupled superconducting systems \cite{goto2021chaos,Berke2022,Borner2024,Cohen2023,benha2025kerr}.  In Ref.~\cite{benha2025kerr} instead it is shown that in present-day Kerr-cat implementations, tunneling is dominated by multimode and multiphoton resonances.   
Their mechanism is \emph{extrinsic}, arising from buffer-mode hybridization.

In this Letter, we identify an \emph{intrinsic} limit: even for an ideal single-mode Kerr-cat, once the phase space becomes mixed chaos-assisted tunneling (CAT) \cite{tomsovic1994cat} introduces a subtle but unavoidable mechanism that degrades the protection of the KCQ. When the classical phase space becomes mixed, regular islands supporting the cat states are surrounded by chaotic regions. Quantum states localized in one island can weakly couple to chaotic states that extend across phase space, effectively bridging both wells and enhancing tunneling by several orders of magnitude.
{Importantly, chaos-assisted tunneling is qualitatively distinct from
previously studied bit-flip channels. Thermal activation requires coupling to
an environment, whereas CAT arises in a fully isolated Hamiltonian. Likewise,
ordinary coherent tunneling involves only a few regular excited states, while
CAT appears only once mixed regular-chaotic phase space develops, enabling
hybridization with a dense chaotic manifold. The resulting enhancement of
splittings is therefore not mere drive-induced leakage into higher-lying
states, but tunneling mediated by chaotic hybridization near the separatrix.}

By combining Floquet simulations with semiclassical WKB \cite{wentzel,brillouin,kramers} and Fermi golden rule estimates, we show that the quasienergy splitting -- and thus the tunneling rate -- exhibits a clear transition as chaos develops. While the static effective model predicts vanishing splittings, the full driven dynamics displays a sharp increase mediated by chaotic states. This constitutes the first quantitative evidence of CAT in a superconducting qubit, identifying chaos as a fundamental mechanism that limits the protection of cat code architectures.

The KCQ is encoded in the eigenstates of the  Kerr parametric oscillator expressed in the rotating frame as
$\hat{H}/\hbar=-K\hat{a}^{\dagger 2}\hat{a}^2+\epsilon_2(\hat{a}^{\dagger 2}+\hat{a}^2)$, where $K$ is the Kerr coefficient, and $\epsilon_2$ the strength of the squeezing interaction (from now on we set $\hbar=1$). 
The Hamiltonian can be written in a factorized form where it is evident that the coherent states $\ket{\alpha=\pm \sqrt{\epsilon_2/K}}$ are degenerate eigenstates of the Hamiltonian. This is the key to the protected qubit encoding\cite{venkatraman2024drive,venkatraman2025thesis}. Although the factorization of the Hamiltonian is an asset that helps towards hardware protection, it is not fundamental. In fact, adding detuning 
\begin{equation}
\label{eq:Hdelta}
\hat{H}_{\rm eff} =\Delta \hat{a}^{\dagger} \hat{a}-K \hat{a}^{\dagger 2} \hat{a}^2+\epsilon_2\left(\hat{a}^{\dagger 2}+\hat{a}^2\right)
\end{equation}
can even improve protection with the appearance of tunable parity-protected degeneracies that occur not only in the ground state manifold, but also in the excited state manifolds of our system \cite{venkatraman2024drive}. In the classical limit, the metapotential defined by {$\hat{H}_{\rm eff}$} has three different phases in the space of parameter $(\epsilon_2/K,\Delta/K)$ is divided by phase transitions at {$\Delta=\pm 2\epsilon_2$}. Here we  consider the case {$-2\epsilon_2<\Delta<2\epsilon_2$}. The corresponding classical Hamiltonian in this case defines a metapotential with a double well structure.
{For an intuitive illustration of the metapotential geometry and the
Bloch-sphere representation of the Kerr-cat encoding, we refer the reader to the Supplemental Material \cite{suppmat}.}
It is integrable with three stationary points. Two of them are stable points at  $\{ \pm \sqrt{(\Delta+2 \epsilon_2)/K}, 0\}$ , which are global minima (bottom of the wells), and $\{0,0\}$ is an unstable hyperbolic point (top of the barrier). The separatrix defining the stable and unstable manifolds is a Bernoulli lemniscate \cite{wielinga1993lemniscate}.

In this regime, the fundamental and first excited eigenstates of the Hamiltonian can be thought of as states 
in a 
spatially separated double-well system.
This separation protects the {KCQ} against bit-flip errors that decay exponentially, while the occurrence of phase-flip errors increases linearly \cite{venkatraman2024drive}. 
These two states are nearly degenerate and can be used to encode quantum information. 

The Hamiltonian \eqref{eq:Hdelta} is the static effective approximation of 
\begin{equation} \label{eq:1}
    \frac{\hat{H}(t)}{\hbar} = \omega_0 \hat{a}^{\dagger} \hat{a} + \sum_{m=3}^{4} g_m (\hat{a}^{\dagger} + \hat{a})^m 
    - i \Omega_d (\hat{a} - \hat{a}^{\dagger}) \cos(\omega_d t)
\end{equation}
where $\omega_0$ and $\omega_d$ correspond to the bare oscillator and the external drive frequencies respectively,
and $g_m$ are third- and fourth-rank nonlinearities \cite{frattini2024observation,venkatraman2024drive}.
It can be implemented experimentally by a superconducting sonlinear asymmetric inductive element (SNAIL) transmon, which consists of a  Josephson junction further shunted by an array of Josephson junctions. Equation~(\ref{eq:1}) is a Taylor expansion about the potential minimum of the SNAIL transmon \cite{venkatraman2025thesis}.
The relation between Eqs.~\eqref{eq:Hdelta} and \eqref{eq:1} is obtained by applying  a displacement into the linear response of the oscillator, where the amplitude of the displacement is $\Pi=\Omega_d\omega_d/\omega_d^2-\omega_0^2$. Then we move into a rotating frame generated by $\omega_d \hat{a}^{\dagger}\hat{a}/2$.\cite{effalg}. The propagator over one period $T$ can be related to the static one by a unitary transformation $\hat{U}=e^{-i\hat{S}/\hbar}$, where $\hat{S}$ is related to micromotion and can be computed perturbatively to arbitrary precision\cite{effalg}.
{In our numerical sweeps we vary the nonlinear coefficient $K$ by tuning
$g_3$ (keeping $g_4$ small), while retuning the drive frequency and amplitude
so that the ratios $\Delta/K$ and $\epsilon_2/K$ remain fixed. Since
$\omega_d$ and $\Omega_d$ are independent experimental controls in SNAIL
devices, this protocol preserves the metapotential shape and isolates the
effect of increasing nonlinearity. Furthermore, in all our calculations we consider  quantities in units of $\omega_0$ (i.e. $K\equiv K/\omega_0$ or $\omega_0=1$)}

To study the driven system, we use
Floquet theory\cite{shirley1965}. The $k$-th eigenstate of the system can be written as $\ket{\psi_k}(t)=e^{-i\varepsilon_k t}\ket{\phi(t)}$, where $\ket{\phi}(t)=\ket{\phi(t+T)}$ are the Floquet modes, with $T$ the period of the drive and $\varepsilon_k$  the Floquet quasienergies. We will operate in the period doubling regime\cite{goto2016bifurcation} so we consider the time evolution map at twice the period of the drive $\tau=2T$. Thus, we have $\hat{U}(\tau)\ket{\phi_k}=e^{-\varepsilon_k \tau}\ket{\phi_k}$. The quasienergies, which are uniquely defined modulo $\omega_d/2=2\pi/\tau$, are obtained by diagonalizing $\hat{U}(\tau)$.

\begin{figure}[ht]
\includegraphics[width=0.95\linewidth]{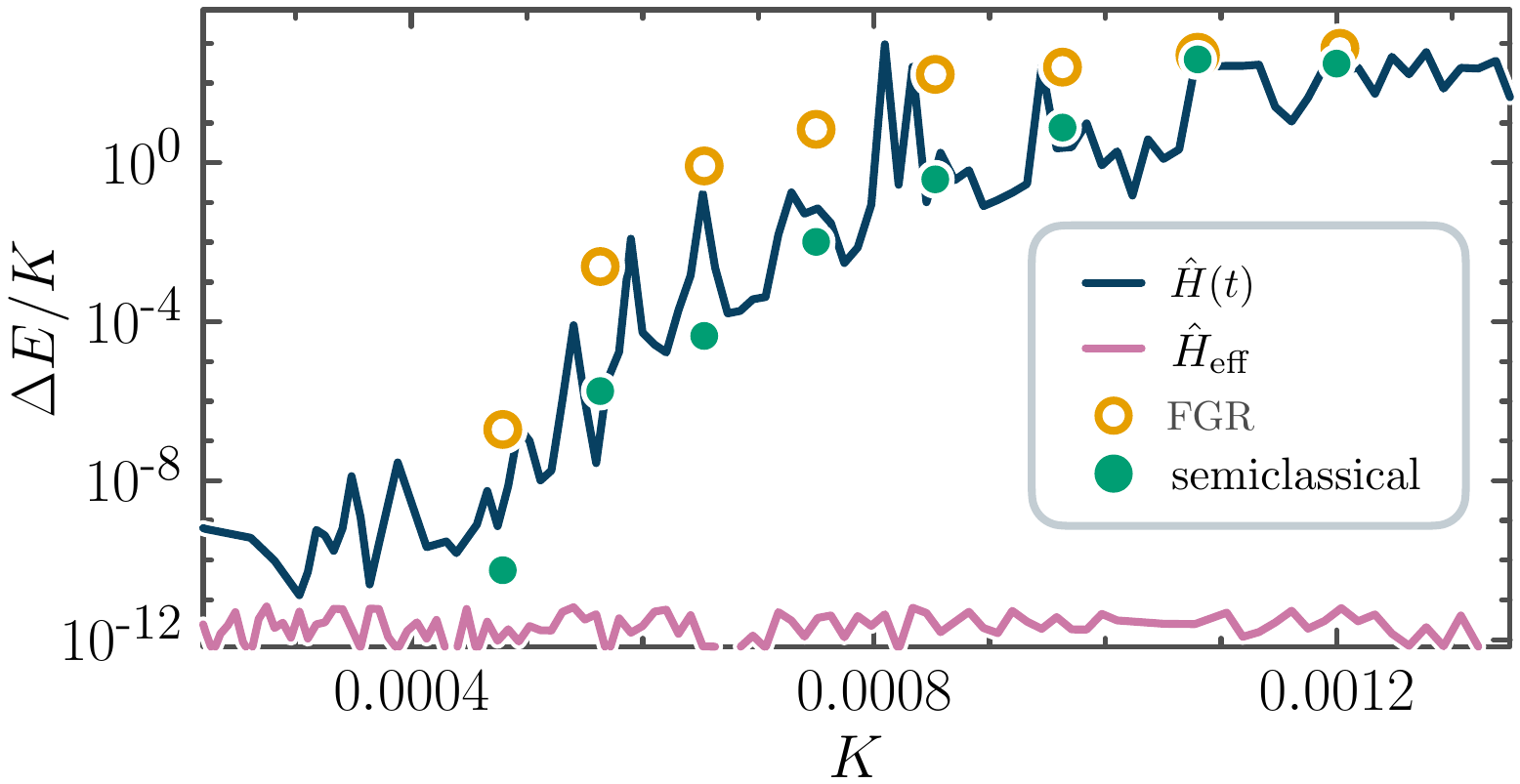}
    \caption{ {Scaled quasienergy splitting $\Delta E/K$ as a function of $K$, for the case $\epsilon_2/K=50$ and $\Delta/K=10$. $N=250$, $g_4=10^{-8}$. The solid blue line corresponds to $\Delta E/K$ computed from $\hat{H}(t)$. The pink (almost constant $\approx 10^{-12}$) line corresponds to $\Delta E/K$ obtained from $\hat{H}_{\rm eff}$. Open circles correspond to quantum calculations. Filled circles correspond to the semiclassical approximation.}
    \label{fig:principal}}
\end{figure}

Under a wide range of parameters, the eigenstates and energies of the static effective Hamiltonian and the Floquet states and quasienergies can be directly related \cite{GarciaMata2024effectiveversus}. However, a key aspect of the qubit protection is the number of states that fit inside the wells. This number is proportional $\Delta/(2K)$ and $\epsilon_2/(piK)-1/2$. These parameters (in particular $\epsilon_2$) are related to nonlinearities and drive amplitude, so increasing them can lead to the development of chaos. The effect of chaos on the states inside the wells has been parametrically mapped, using localization properties \cite{GarciaMata2024effectiveversus}.  Furthermore,  at the hyperbolic point, there is an excited-state quantum phase transition (ESQPT), the breakup -- due to chaos -- of which hinders the potential for stabilized cat qubits\cite{ garciamata2025impact}.

Aside from affecting the correct description by a static Hamiltonian, chaos can have a direct effect on the qubit stability. This stability is favored by the suppression of tunneling, observed by $\Delta E=0$ for $\Delta/K=2 m$ (with $m$ integer) in $\hat{H}_{\rm eff}$ of $\eqref{eq:1}$ \cite{venkatraman2024drive}. 
In Fig.~\ref{fig:principal} we compute $\Delta E=E_1-E_0$, for a fixed $\Delta/K=10$ and $\epsilon_2/K=50$, from $\hat{H}_{\rm eff}$ (pink line), and $[(\epsilon_1-\epsilon_0)\mod (\omega_d/2)]/K$ from the 
 Floquet operator $\hat{U}(2\tau)$ (blue line), where $\epsilon_{0,1}$ are quasienergies and their order can be identified  using an iterative procedure \cite{GarciaMata2024effectiveversus}.  
We observe that the splitting remains constant and close to $0$ ($10^{-12}$ within computational precision),  as $K$ increases. This is expected for even $\Delta/K$ \cite{venkatraman2024drive}.
On the other hand, the curve obtained from the time dependent system, begins with a plateau around $\approx 10^{-8}$ set by the numerical error in computing $\ket{U}$, computed using the Adams integration method. At around $K\approx 0.0005$ the splitting increases by several orders of magnitude with $K$, 
before eventually saturating to a plateau.
Below we show that this increase in the splitting is directly related to the presence of chaos.
{For typical circuit frequencies $\omega_0/2\pi = 4$–$8~\mathrm{GHz}$,
the sweep in Fig.~1 corresponds to $K/2\pi \approx 0.4$–$14~\mathrm{MHz}$,
thus spanning both present-day Kerr-cat implementations and the larger
nonlinearities envisioned in next-generation devices. We remark that we have verified that the same qualitative behavior shown in Fig.~\ref{fig:principal} arises when the $K$ is held fixed and $\epsilon_2/K$ is varied; see
Supplemental Material for details \cite{suppmat}.}

\begin{figure}[t]
    \includegraphics[width=0.9\linewidth]{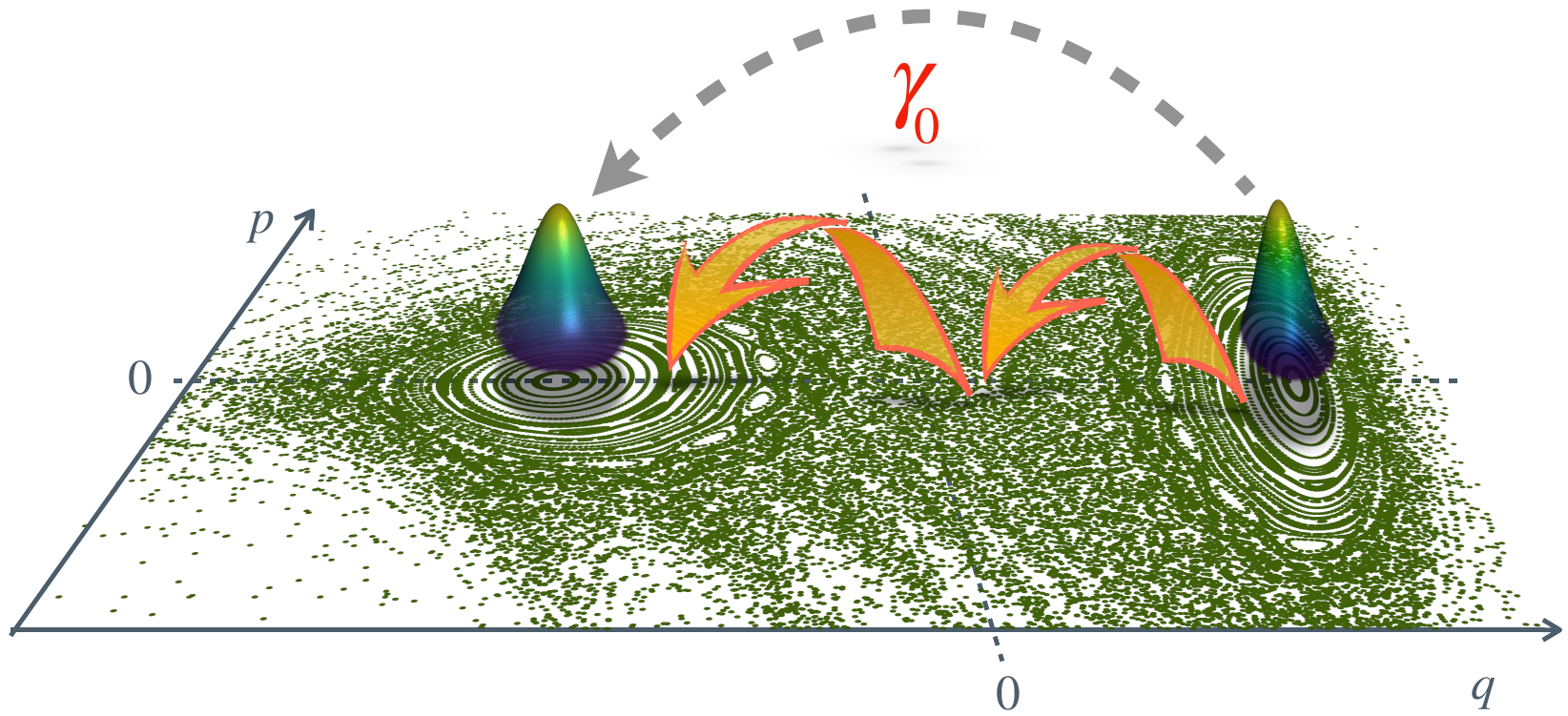}
    \caption{Visual interpretation of the CAT affecting the ground state of the Kerr parametric oscillator. The green points correspond to the classical Poincar\'e section. The tunneling takes place, mediated by a chaotic state with quasienergy near that of the ground state, with a rate $\gamma$.     \label{fig:poincare}}
\end{figure}

The cat qubit is a superposition of coherent states each living in a double well metapotential. These wells are regular islands in phase space. If there is chaos surrounding the wells, the probability tails of states inside one well can overlap with scattered states in the chaotic sea, classically propagating via chaotic diffusion until they are close enough to tunnel into the other well.
This phenomenon is known as chaos assisted tunneling \cite{tomsovic1994cat} and has been experimentally observed 
in billiards (microwave in annular billiards \cite{Dembowski2000First} and light in deformed circular billards \cite{Shinohara2010}), cold atoms \cite{Steck2001,Steck2002,Arnal2020} and microcavity lasers \cite{Kim2013}. In Fig.~\ref{fig:poincare}, we schematically illustrate this tunneling process, where on the plane there is a Poincar\'e section for the actual physical system considered. This effect can be measured using the decay rate $\gamma_0$ from the fundamental state to the chaotic sea and it can be linked with the energy splitting ($\Delta E$).
{Localization of the in-well states is diagnosed using the inverse
participation number, the Wehrl entropy, and inspection of the Husimi
distributions, as detailed in the Supplemental Material \cite{suppmat}.}

To understand this difference between the energy splittings and further link the phenomenon with chaos, we follow previous studies \cite{tomsovic1994cat,backer2008regular} and calculate the tunneling rate and energy splitting of a mixed system via quantum calculations. Here, we take advantage of the effective description to define a fictitious integrable system that resembles the dynamics of the regular Floquet eigenstates as closely as possible but also extending the regular dynamics beyond the integrable part, and with it get a semiquantumantum calculation of the tunneling rate.

As shown in \cite{backer2008regular}, we can obtain the decay rate $\gamma_0$ of the fundamental eigenstate to the chaotic sea using Fermi golden rule (FGR) and assuming that the system is a mixed system, i.e the chaotic behavior can be expressed as a perturbation to a regular quartic Hamiltonian. At this point, we consider that the Kolmogorov-Arnold-Moser resonances \cite{Lichtenberg1992} that might be present for near-integrable tori have an area $A\ll\hbar\approx 1$ rendering them invisible in the quantum regime, which seems to be acceptable for the parameters taken and the eigenstates observed throughout the simulations. This ensures that resonance-assisted tunneling (RAT)\cite{keshavamurthy2007dynamical} is not present in our system and that the increase in quasienergy difference is only due to CAT. 
To define a fictitious integrable system, the Hamiltonian is decomposed into regular and chaotic parts using projectors into the basis of the effective Hamiltonian, as introduced in Ref.~\cite{Podolsky},
\begin{equation}
\hat{H}
= E_{R}\,\hat{P}_{\mathrm{reg}}
+ E_{C}\,\hat{P}_{\mathrm{ch}}
+ \sum_{m,n} \left( V_{mn}\, |\psi^{\mathrm{reg}}_{m}\rangle \langle \psi^{\mathrm{ch}}_{n}| + \text{h.c.} \right) ,
\label{eq:3}
\end{equation}
where the projectors can be defined as
\begin{equation}
    \hat{P}_{\text{reg}} = \sum_{m=1}^{N_{\text{reg}}}\ket{\psi_m^{\text{reg}}} \bra{\psi_m^{\text{reg}}}, \quad
    \hat{P}_{\text{ch}} = \sum_{m=1}^{N_{\text{ch}}}\ket{\psi_m^{\text{ch}}} \bra{\psi_m^{\text{ch}}},
\end{equation}
with $\ket{\psi_m^{\text{ch}}}$ and $\ket{\psi_m^{\text{reg}}}$ eigenstates of the effective Hamiltonian, {and $N_{\text{reg}}$ and $N_{\text{ch}}$ the number of states in each projection}. Indeed, these eigenstates need to be orthogonal and define a complete basis such that $\hat{P}_{\text{reg}}+\hat{P}_{\text{ch}}=\mathbb{I}$. 

The exponential decay $e^{-\gamma_0t}$ from the fundamental state to the chaotic sea has a rate 
\begin{equation}
    \gamma_0=||\hat{P}_{\text{ch}} \hat{U}\ket{\psi_0^{\text{reg}}}||^2.
    \label{eq:5}
\end{equation}
obtained from the coupling matrix element of the time evolution $\hat{U}$ of the system and FGR in a discrete spectrum.

This remains valid if one considers the decay up to the Heisenberg time $t_H=\frac{\hbar}{\Delta_{\text{ch}}}$ where $\Delta_{\text{ch}}$ is the mean level spacing for chaotic states. Here, $\gamma_0$ and $t$ have been adimensionalized to multiples of period $T=\frac{4\pi}{\omega_d}$ and the density of chaotic states used in the discrete FGR is $\rho_{\rm ch}=\frac{1}{\Delta_{\rm ch}}\propto \frac{1}{N_{\rm ch}}$. We remark that to apply this method we have assumed that the tails of probability generated by the one period evolution are responsible for overlapping with chaotic states and inducing a transition probability.
The decay rate then depends on the regular Hamiltonian taken, and therefore it has to bear the same dynamics as closely as possible.

In order to link the decay rate with the measurements of quasienergy splitting of the fundamental and first excited state ($\Delta E_0$), we follow the idea of \cite{backer2008regular} and \cite{tomsovic1994cat} where it is established that the increase in splitting is due to the presence of nearby chaotic states whose distance to regular states fluctuates. According to random matrix models, $\Delta E_0$ follows a Cauchy distribution with a defined geometric mean, that coupled to Eq.~\ref{eq:5} gives us the simple relation to $\gamma_0$
\begin{equation}
    \langle\Delta E_0\rangle=\gamma_0.
    \label{eq:6}
\end{equation}
The main challenge in this method then lies in identifying the chaotic states, because they determine the extent of the chaotic sea in the time-independent basis. Classical tools such as Lyapunov exponents \cite{Benettin} can be used to separate chaotic and regular eigenstates by determining the last invariant orbit and Einstein-Brillouin-Keller quantization \cite{Benettin} to count the number of regular states below it. 
Alternatively, localization in phase space can be used 
to characterize chaotic eigenstates,
as they typically appear spread over the chaotic sea in any quasi-probability distribution. 
We use the inverse participation number and the Wehrl entropy computed with the Husimi Q distribution for a given effective eigenstate and compare that to the corresponding Floquet quasi-state (see Supplemental Material \cite{suppmat}). 

We show the results obtained with this method in 
Fig.~\ref{fig:principal} (orange border circles). They agree well with the Floquet simulations, displaying a sharp initial increase that eventually saturates at a plateau, where the quasi-probability distribution is completely delocalized.
Remarkably, even when the difference between the system and its approximation spans several orders of magnitude, the quantum states remain structured and localized. This indicates that the effects of chaos emerge before the erasure of the encoding states, inducing tunneling between islands, i.e. bit-flip errors. Hence, chaos enhances such errors, which can occur without visible delocalization.  This method performs reliably for all tested values of $\tfrac{\epsilon_2}{K}$ and $\tfrac{\Delta}{K}$ (see Supplemental Material\cite{suppmat}).
These quantum calculations provide evidence of CAT in the KCQ. 
{The emergence of enhanced splittings while the logical states remain
localized demonstrates that CAT can induce bit-flip errors even before global
delocalization sets in, revealing an intrinsic dynamical limitation that is
absent in the static effective description.}

To extend our analysis even further,
we employ a different technique as the semiclassical limit is approached, i.e. when a large number of states populate the wells.  A universal semiclassical method for chaos-assisted tunneling was developed in Ref.~\cite{Podolsky}.  Assuming that regular islands are  surrounded by a structureless chaotic sea and the absence of quantifiable KAM resonances, one can apply semiclassical quantization of in-well states (e.g. by WKB approximation) to calculate the variance of the coupling matrix elements between regular states and the chaotic sea. Following Refs.~\cite{Podolsky,keshavamurthy2007dynamical}, we obtain an energy-splitting formula
\begin{equation}
    \Delta E = c_0\hbar \, 
\frac{\Gamma\!\left(\tfrac{A}{\pi \hbar}, \, 2 \tfrac{A}{\pi \hbar}\right)}
     {\Gamma\!\left(\tfrac{A}{\pi \hbar} + 1, \, 0\right)},
     \label{eq: 7}
\end{equation}
where $\Gamma$ is the incomplete gamma function, $c_0$ is a proportionality factor that depends on the system and its parameters, and $A$ is the area of the regular islands. 

Results for this method are shown as blue circles in Fig.~\ref{fig:principal}. They follow the quasienergy splitting closely and reproduce the behavior of the simulations, including the plateau reached at higher values of $K$. This shows further evidence of the effect of CAT phenomenon in the KCQ, now in a semiclassical calculation as it is usual in chaotic phenomena. The choice of parameters in this figure allow for many states to be enclosed in the wells, enough to agree with calculations in the semiclassical limit. In the Supplemental material \cite{suppmat} we show that discrepancies are found for lower values of $\frac{\epsilon_2}{K}$ and $\frac{\Delta}{K}$. However, in this regimes, quantum calculation excel at reproducing the simulation data.

In summary, we have demonstrated that chaos-assisted tunneling imposes a fundamental limit to the protection of the Kerr-cat qubit. While the static effective Hamiltonian predicts exponentially suppressed tunneling between the cat states, the full driven dynamics reveals that chaotic states mediate tunneling across the double-well metapotential, producing large quasienergy splittings.

By combining Floquet simulations, fully quantum decay-rate calculations, and semiclassical WKB estimates, we have quantitatively established that the tunneling rate increases sharply at the onset of chaos and saturates once chaotic mixing becomes complete. The agreement between these independent approaches provides the first direct evidence of CAT in a superconducting qubit.
{At the onset of mixed dynamics, the quasienergy splitting for
representative stabilized-cat parameters reaches the Hz scale, corresponding
to CAT-limited bit-flip times of order $0.1$--$1~\mathrm{s}$ and thus
providing a quantitative lower bound on the coherence attainable as $K$ is
increased.}
These results identify chaos as an intrinsic error channel that emerges even before global delocalization of the logical states. In practical terms, chaos-enhanced tunneling sets a bound on the coherence times achievable in dynamically protected qubits, constraining the parameter space available for scalable cat code architectures.
{While our quantitative analysis is specific to the voltage-driven SNAIL implementation, the underlying mechanism—suppression and subsequent chaos-assisted enhancement of tunneling in a driven nonlinear double well—may also arise in other Kerr-cat architectures. Whenever the classical phase space features a saddle separating two stable minima, increasing the drive generically breaks the separatrix and produces mixed regular–chaotic dynamics, suggesting that analogous CAT behavior could appear in designs such as the flux-driven symmetrically threaded SQUID\cite{STS2025}.}
More broadly, this work highlights that the interplay between nonlinear dynamics, periodic driving, and chaos can fundamentally limit the coherence of quantum hardware, establishing a new frontier at the intersection of quantum chaos and fault-tolerant quantum computation.

\begin{acknowledgments}
\textit{Acknowledgments--}
We acknowledge support from Argentinian Agencia I$+$D$+$i (Grants No. PICT-2020-SERIEA-00740 and PICT-2020-SERIEA-01082). I.G.-M. received support from the French-Argentinian International Research Project \textit{Complex Quantum Systems} (COQSYS), funded by CNRS.
D.A.W.  received support from CONICET (Grant No. PIP 11220200100568CO), UBACyT (Grant No. 20020220300049BA) and PREI-UNAM.
\end{acknowledgments}

%

\onecolumngrid

\clearpage
\newpage
\renewcommand{\thefigure}{S\arabic{figure}}
\renewcommand{\thetable}{S\arabic{table}}
\renewcommand{\theequation}{S\arabic{equation}}
\renewcommand{\thesection}{S\arabic{section}}
\setcounter{figure}{0}
\setcounter{table}{0}
\setcounter{equation}{0}
\setcounter{section}{0}
\renewcommand{\thefigure}{S\arabic{figure}}
\renewcommand{\thetable}{S\arabic{table}}
\renewcommand{\theequation}{S\arabic{equation}}
\renewcommand{\thesection}{S\arabic{section}}

\begin{center}
\textbf{    {\large Supplemetary material for ``Fundamental Limits to Cat Code Qubits from Chaos-Assisted Tunneling''}}

\vskip 1cm

Lionel E. Mart\'inez,$^{1,2}$, Ignacio Garc\'ia-Mata,$^3$ and Diego A. Wisniacki$^1$

\textit{
$^1$Departamento de F\'isica ``J. J. Giambiagi'' and IFIBA, FCEyN,\\
Universidad de Buenos Aires, 1428 Buenos Aires, Argentina}

\textit{ $^2$ Department of Physics \& Astronomy, University of Southern California, Los Angeles, CA 90089, United States of America}

\textit{$^3$ Instituto de Investigaciones F\'isicas de Mar del Plata (IFIMAR),\\
FCEN, Universidad Nacional de Mar del Plata\\ 
\& CONICET, Funes 3350 (7600) Mar del Plata, Argentina
 }

\end{center}

This Supplemental Material discusses (i) further results for different values of the detuning parameter ($\frac{\Delta}{K}$) and $\frac{\epsilon_2}{K}$, and (ii) the methods for the selection of the chaotic projector states.

\subsection{Metapotential and Bloch-sphere picture of the Kerr-cat qubit}
\label{sec:metapot_bloch}

\begin{figure}[h]
    \centering
    \includegraphics[width=0.95\linewidth]{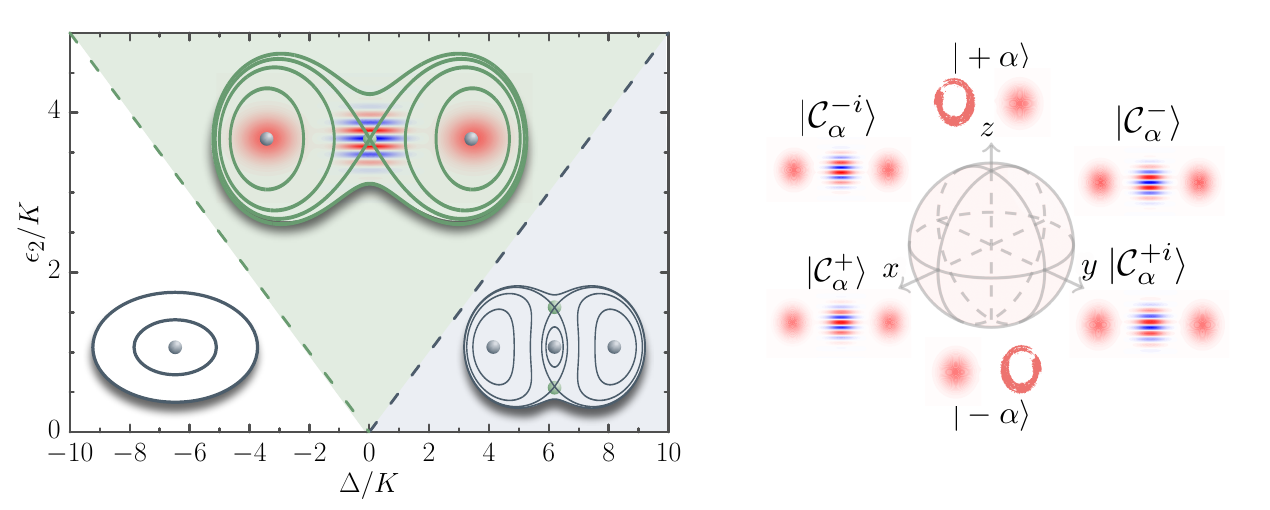}
    \caption{{Left: schematic phase diagram of the classical metapotential of
    the detuned Kerr parametric oscillator in the $(\Delta/K,\epsilon_2/K)$ plane.
    The shaded wedge $|\Delta| < 2\epsilon_2$ corresponds to the double-well regime
    of the effective Hamiltonian~(1) of the main text, where the metapotential
    exhibits two symmetry-related minima separated by a barrier (central panel). On top of the metapotential lines, we show the Wigner function of the ground state for this case.
    Outside this region the metapotential has a single minimum (bottom-left panel)
    or a more complicated structure (bottom-right panel). Right: Bloch-sphere
    representation of the logical manifold spanned by coherent states
    $\ket{\pm\alpha}$ and their even/odd superpositions. Along the $z$ axis we
    place the coherent states $\ket{+\alpha}$ and $\ket{-\alpha}$, while the
    equatorial directions correspond to cat states
    $\ket{C_\alpha^{\pm}} \propto \ket{\alpha}\pm\ket{-\alpha}$ and their
    $\pi/2$-phase-rotated counterparts
    $\ket{C_\alpha^{\pm i}}\propto \ket{i\alpha}\pm\ket{-i\alpha}$. This provides
    a convenient visualization of logical $Z$ (motion along the meridians) and
    logical $X/Y$ (rotations on the equator) operations within the protected
    subspace.}}
    \label{fig:metapot_bloch}
\end{figure}

{
In this section we provide an intuitive illustration of the metapotential of the Kerr
parametric oscillator and of the logical encoding of the Kerr-cat qubit on an effective
Bloch sphere, as shown in Fig.~\ref{fig:metapot_bloch}. This material is intended as a
visual complement to the main text, and does not introduce any additional results.}

{
The effective metapotential associated with the Hamiltonian
$\hat{H}_{\mathrm{eff}} = \Delta \hat{a}^\dagger \hat{a}
- K \hat{a}^{\dagger 2}\hat{a}^2
+ \epsilon_2(\hat{a}^{\dagger 2} + \hat{a}^2)$
depends only on the dimensionless ratios $\Delta/K$ and $\epsilon_2/K$.
In the classical limit, this potential displays three distinct regimes in the
$(\Delta/K,\epsilon_2/K)$ plane. For $|\Delta| < 2\epsilon_2$ the metapotential
has a double-well structure with two symmetry-related minima separated by a
barrier located at the origin. In this regime the phase space contains two large
regular islands around the minima, separated by the separatrix associated with
the hyperbolic fixed point at the barrier top. This is the regime relevant for
the Kerr-cat qubit, as discussed in the main text. We have added a Wigner function representation of the ground state, which is seen to have large (Gaussian-like) contributions sitting on the well bottoms, and the interference fringes in the middle. For $|\Delta| > 2\epsilon_2$,
the metapotential exhibits a single minimum and the double-well structure is
lost.}

{
In the double-well regime, the logical manifold of the Kerr-cat qubit can be
visualized as an effective Bloch sphere embedded in the infinite-dimensional
Hilbert space. In Fig.~S1, we show the basis states represented by their Wigner function\cite{wignerfunction}. A natural choice of poles is given by the coherent states
$\ket{\pm\alpha}$ localized at the two minima of the metapotential, with
$\alpha \simeq \sqrt{\epsilon_2/K}$ in the static effective description.
Even and odd cat states
$\ket{C_\alpha^{\pm}} \propto \ket{\alpha} \pm \ket{-\alpha}$
then form approximate eigenstates of the effective Hamiltonian and can be
associated with the logical $\ket{0_L}$ and $\ket{1_L}$ states along one
equatorial axis. Phase-rotated cats
$\ket{C_\alpha^{\pm i}} \propto \ket{i\alpha} \pm \ket{-i\alpha}$
span the orthogonal equatorial axis. In this representation, logical $Z$
operations correspond to relative phase accumulation between the wells
(i.e.\ motion along meridians connecting $\ket{+\alpha}$ and $\ket{-\alpha}$),
while logical $X$ and $Y$ rotations are implemented as transformations within
the cat-state manifold on the equator.

This Bloch-sphere picture will be useful for interpreting how chaos-assisted
tunneling affects the logical manifold: CAT effectively opens a leakage channel
between the two wells through chaotic states living outside the logical
subspace, thereby inducing bit-flip errors even when the cat states themselves
remain sharply localized in phase space.}

\subsection{Further results}

\begin{figure}[htbp]
\includegraphics[width=.8\linewidth]{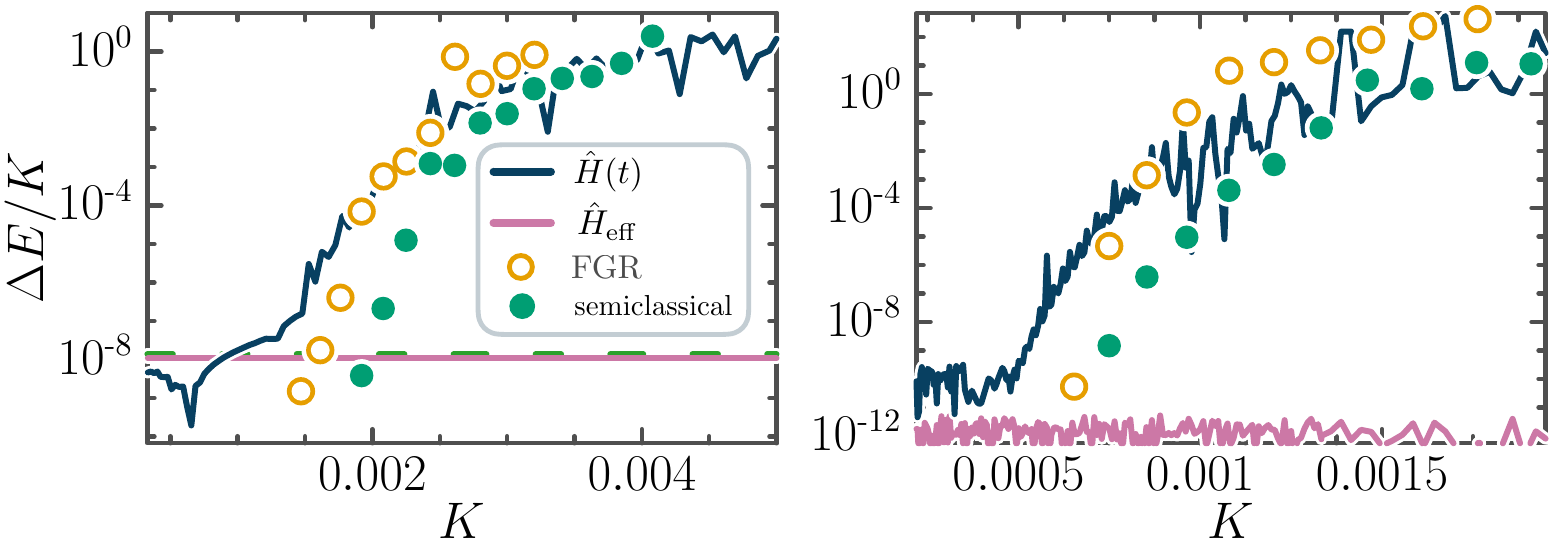} 
\caption{Scaled quasi-energy $\Delta E/K$ as a function of $\rm K$, for the cases $\epsilon_2/K=10$ and $\Delta/K=0.2$ (left panel), and $\epsilon_2/K=30$ and $\Delta/K=10$ (right panel). $N=250$, $g_4=10^{-8}$. Solid blue lines for both panels
correspond to $\Delta E/K$ computed from the time-dependent Hamiltonian. The pink
line corresponds to $\Delta E/K$ obtained for the static effective Hamiltonian with the dashed green line on the left panel exhibiting the splitting limit shown in \cite{venkatraman2025thesis}. Open circles correspond to quantum calculation. Filled circles correspond to the semiclassical approximation.}
\label{fig:1S}
\end{figure}

Figure~1 of the main text shows the energy splitting between the ground and first excited states, together with the quantum and semiclassical calculations for the decay rate of these states into the chaotic sea, which coincides with the energy difference. The system parameters for this figure are $\frac{\epsilon_2}{K}=50$ and $\frac{\Delta}{K}=10$, allowing for approximately 17 states -- sufficient to demonstrate a reliable semiclassical approximation (blue circles). In this Supp. Mat. we present additional results  (see Fig. \ref{fig:1S}), where we explore lower parameter values and find excellent agreement between the Floquet energy difference and the quantum and {semiclassical calculations of Eqs.~(6) and (7).}

We begin with the left panel, which corresponds to the lowest parameter values, $\Delta/K=0.2$ and $\epsilon_2/K=10$. In this regime, a quasi-energy difference above the computational threshold near $\Delta/K = 0$ allows for a more detailed analysis and shows the presence of detunning shifts on the time-dependent Hamiltonian.  The effective energy splitting (pink line) remains constant across all values of $K$, with no fluctuations. This stability arises because the chosen parameters render the first pair of states quasi-degenerate, with a splitting several orders of magnitude larger than numerical errors. The black dashed line marks the maximum possible effective splitting at $\epsilon_2/K=10$, whose expression was derived in \cite{venkatraman2025thesis}. This bound provides a useful indicator of the onset of chaotic behavior: any observed energy difference exceeding it must result from nonlinear effects, not present in the effective approximation. The Floquet results (blue line), unlike in Fig.~1 (main text), show both the system and its effective approximation coincide for low values of the nonlinear parameter. The sudden, regular dip that comes after in the Floquet splitting reflects energy shifts induced by $K$, and Stark and Lamb shifts, which were also observed and discussed experimentally in \cite{venkatraman2024drive}. Remarkably, the quasi-energy splitting remains regular up until the  energies upper bound, beyond that the energy difference begins to rise dramatically and shows fluctuations that are chaotic in nature. This is further evidence that the increase in the tunneling rates that induce bit-flip errors are due to chaotic behavior. 

We can also see that after this bound the quasi-energy difference is effectively described by the quantum calculations (orange circles), showing excellent agreement with simulations. Instead, the semiclassical approximation (blue circles) fails to accurately describe the decay rate outside of the plateau reached for higher values of $K$, expected given we are far away from the semiclassical regime.  

The right panel shows an intermediate parameter between the figure in the left panel and Fig.~1 (main text). A similar behavior of the energy splitting is observed, with a sudden increase in the energy gap between the fundamental and first excited states. The quantum calculations (orange circles) continue to show excellent agreement with the simulated system splitting, while the semiclassical calculations tend to be closer to the Floquet splittings than for lower parameter values although still being several orders of magnitude under the expected value.

We have therefore demonstrated a wide range of validity for our calculations, finding evidence of chaos-assisted tunneling in the KCQ, which is expected to impact its performance at high nonlinear parameters due to bit-flip errors.
\subsection{Robustness of chaos-assisted tunneling under alternative parameter sweeps}
\label{sec:epssweep}
\begin{figure}[htbp]
\includegraphics[width=.4\linewidth]{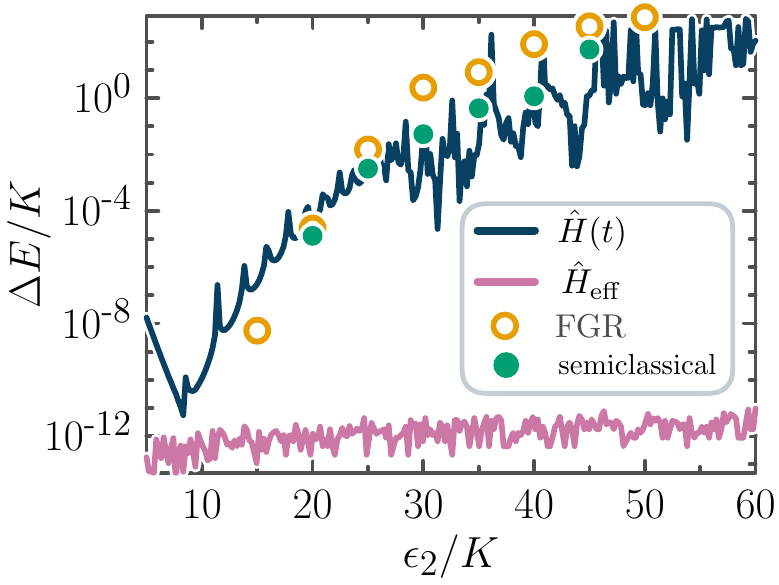} 
\caption{
{Scaled quasi-energy splitting $\Delta E/K$ obtained from Floquet simulations
as a function of the drive strength $\epsilon_2/K$, for fixed Kerr
nonlinearity $K=0.00102$ and detuning $\Delta/K=10$. $N=250$, $g_3=0.0175$ and $g_4=10^-8$.
}}
\label{fig:epssweep}
\end{figure}

{
In Fig.~1 of the main text, we  explore the emergence of chaos-assisted tunneling
(CAT) by sweeping the Kerr nonlinearity \(K\) while keeping the ratios
\(\Delta/K\) and \(\epsilon_2/K\) fixed. Since experimentally one may also vary
the two-photon drive strength at fixed Kerr nonlinearity, here we demonstrate
that the CAT phenomenology is not specific to this choice of sweep.

In Figure~\ref{fig:epssweep} we show the quasi-energy splitting \(\Delta E/K\)
obtained from the simulations when \(K\) is held fixed ($K=0.00102$) and the ratio
\(\epsilon_2/K\) is varied by varying $\Omega_d$. The remaining parameters are $\Delta/K=10$, $g_3=0.0175$ and $g_4=10^{-8}$. As \(\epsilon_2/K\) is increased, we observe
the same qualitative behavior: an initial regime where the splitting remains
exponentially suppressed, followed by a sharp increase over several orders of
magnitude, and eventual saturation at a plateau once the dynamics becomes fully
mixed.

This behavior mirrors the results obtained under a \(K\)-sweep and confirms
that the enhancement of tunneling is controlled by the onset of mixed
regular–chaotic phase space rather than by the specific microscopic parameter
being varied. Physically, increasing \(\epsilon_2/K\) enlarges the number of
in-well states and strengthens nonlinear resonances, which leads to the
breakup of the separatrix and the formation of a chaotic sea that mediates
tunneling between the two regular islands.

We therefore conclude that chaos-assisted tunneling is a robust and generic
feature of the driven Kerr-cat Hamiltonian, arising whenever system parameters
are tuned into a regime of mixed classical dynamics, independently of whether
this is achieved by increasing the Kerr nonlinearity or the effective drive
strength.}
\subsection{Quantum calculations}
The calculation of the energy splittings associated with Eq.~(5) [main text] requires identifying the chaotic states of the system in order to construct a chaotic projector from the effective eigenstates. The size of this chaotic subspace determines the overlap with the fundamental eigenstate, and thus its decay rate into the sea. In analogy with the classical ergodic exploration of the chaotic sea, quantum chaotic eigenstates delocalize and hybridize throughout the sea. This delocalization provides a way to estimate the extent of the chaotic sea by comparing the eigenstates with their effective counterparts, while keeping the calculation fully quantum mechanical.

To quantify this delocalization, we define the inverse participation number (IPN) using the Husimi-Q distribution of a given eigenstate and compare it with the corresponding IPN computed from the Floquet quasi-states. This comparison is essential, since there is no absolute reference for localization in the absence of a fully regular system. Moreover, since the quasi-energy states are not ordered, each must be matched to a regular eigenstate by identifying the one with the highest fidelity overlap. If one quasi-state is not successfully matched, then it has to belong to the chaotic sea. The IPN calculation for a Husimi distribution follows
\begin{equation}
    \text{IPN}=\frac{1}{\sum_n Q(x_n,p_n)\Delta x \Delta p},
\end{equation}

here $Q(x_n,p_n)$ is the discrete value of the Husimi distribution at the point $(x_n,p_n)$. The distribution is discrete since it is numerically calculated. We conclude that when the IPN of the effective Hamiltonian differs consistently from the corresponding Floquet measurements, we are at the beginning of the chaotic sea. This chaotic sea might separate the left and right invariant tori or be outside the lemniscate. Generally speaking, the delocalization is appreciable and easily identified in the quasi-state Husimi distribution plots, which were observed to further confirm the extent of the chaotic sea.

\begin{figure}[htbp]
\includegraphics[width=.8\linewidth]{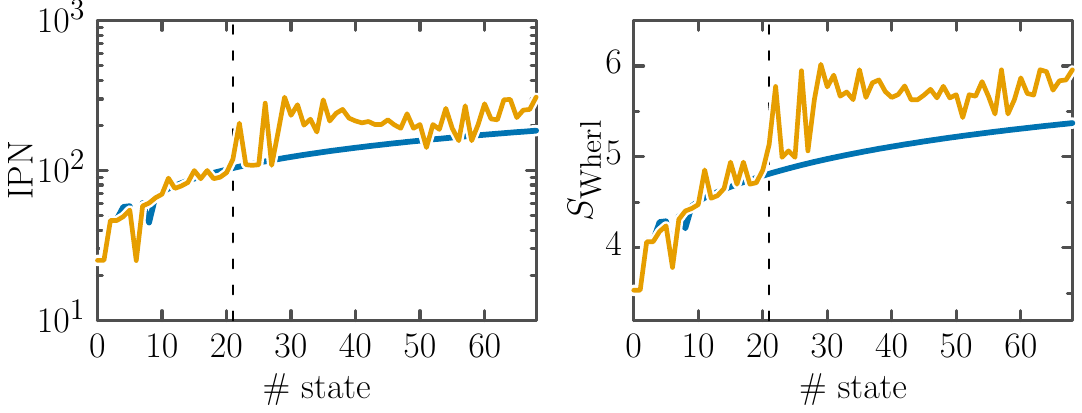} 
        \caption{IPN (left panel) and Wehrl entropy (right panel) comparison between the effective approximation (blue line) and the Floquet-solved system (orange line). Parameters are $\Delta/K=0.2$, $\epsilon_2/K=10$, and $K=0.002$. These sets of parameters correspond to a point on the left panel of Fig.~\ref{fig:1S}. Both quantities serve as indicators of delocalization in the coherent-state basis.}
        \label{fig:2S}
\end{figure}

An example of the identification of the chaotic sea via IPN calculation is shown in Fig.~\ref{fig:2S}(Left) for the parameters $\frac{\Delta}{K}=0.2$, $\frac{\epsilon_2}{K}=10$ and $K=0.002$ corresponding to the Fig.~\ref{fig:1S}(Left). Here a consistent, clear difference in the localization of regular eigenstates and quasi-states is identified and marked with a vertical line, allowing us to define the start of the chaotic sea marked by a vertical dashed line. 

The consistent difference must be met since classically, the nonlinearity breaks regular orbits inwards. Then, any state that is considered chaotic must be followed by higher energy chaotic states. We may see isolated sharp difference in the plots but those are usually not predominant and arise from state identification errors.

We go further and add another measure of delocalization as a complement: the Wehrl entropy. It provides a phase-space characterization of eigenstate spreading by quantifying the information content of the Husimi distribution. In this context, low Wehrl entropy signals localization and corresponds to regular eigenstates, while high entropy reflects delocalization and is therefore a marker of chaotic behavior when compared to an effective approximation, useful to verify the extent of the chaotic sea obtained from other measures. The Wehrl entropy is defined as
\begin{equation}
    S_{\text{Wehrl}} 
= - \sum_i Q(x_i, p_i)\,\ln\!\big(Q(x_i, p_i)\big)\,\Delta x \Delta p .
\end{equation}
where $Q(x_n,p_n)$ is the discrete value of the Husimi distribution at the point $(x_n,p_n)$.

Results of this type of measurements are shown and compared against IPN results in Fig.~\ref{fig:2S}, where again a clear difference between regular eigenstates and Floquet quasi-states can be determined marking the chaotic sea. These measurements are repeated for each blue point in Figs.~1(main text)  and \ref{fig:1S} and are essential to determine the extent of the chaotic sea.

\end{document}